\documentclass[preprint,showpacs,preprintnumbers,nofootinbib,amsmath,amssymb]{revtex4}
\usepackage{graphicx}
\usepackage{dcolumn}
\usepackage{bm}

\newcommand{\beq} {\begin{equation}}
\newcommand{\eeq} {\end{equation}    }
\newcommand{\bea} {\begin{eqnarray} }
\newcommand{\eea} {\end{eqnarray}    }

\newcommand{\Lg}{\,{\mathcal L}}
\newcommand{\lm}{\lambda}
\newcommand{\no}{\nonumber}
\newcommand{\gm}{\gamma}
\newcommand{\es}{\epsilon}

\newcommand{\Gm}{\Gamma}
\newcommand{\sg}{\sigma}
\newcommand{\dt}{\delta}
\newcommand{\kp}{\kappa}
\newcommand{\Dt}{\Delta}
\newcommand{\di}{\rm d}
\newcommand{\mw}{m_W}

\newcommand{\sbt}{s_{\beta}}
\newcommand{\cbt}{c_{\beta}}
\newcommand{\tbt}{t_{\beta}}
\newcommand{\shat}{{\hat{s}}}
\newcommand{\that}{{\hat{t}}}
\newcommand{\uhat}{{\hat{u}}}
\newcommand{\bbar}{{\bar{b}}}
\newcommand{\Fs}{{F_{1/2}^H}}
\newcommand{\Fp}{{F_{1/2}^\eta}}

\newcommand{\thin}{\hspace{.1pt}}
\newcommand{\lsim}{\mathrel{\mathop{\kern 0pt \rlap
  {\raise.2ex\hbox{$<$}}}
  \lower.9ex\hbox{\kern-.190em $\sim$}}}
\newcommand{\gsim}{\mathrel{\mathop{\kern 0pt \rlap
  {\raise.2ex\hbox{$>$}}}
  \lower.9ex\hbox{\kern-.190em $\sim$}}}
\begin{document}
\title {Associated production of a single heavy $T$-quark \\
in the littlest  and simplest little Higgs models}
\author{Kingman Cheung$^{1,2}$, C.S. Kim$^{3}$, Kang Young Lee$^4$, and
Jeonghyeon Song$^5$}
\address{
$^1$Department of Physics, National Tsing Hua University,
Hsinchu, Taiwan R.O.C. \\
$^2$National Center for Theoretical Sciences, Hsinchu, Taiwan R.O.C. \\
$^3$Department of Physics, Yonsei University, Seoul 120-749, Korea\\
$^4$Department of Physics, KAIST, Daejeon 305-701, Korea\\
$^5$Department of Physics, Konkuk University, Seoul 143-701, Korea
}
\date{\today}
\begin{abstract}
\noindent The colored $SU(2)_L$-singlet  heavy $T$-quark is one of the
most crucial ingredients in little Higgs models,  which is
introduced to cancel the largest contribution of the SM top quark to
the Higgs boson mass at one-loop level. In two representative little
Higgs models, the littlest Higgs model and the $SU(3)$ simplest
Higgs model, we comprehensively study the single heavy $T$-quark
production at Large Hadron Collider (LHC). After presenting the
possibility of relatively light ($\sim 500$ GeV) $T$-quark in the
simplest little Higgs model, we consider all the relevant processes,
the $2 \to 2$ process of $q b \to q' T$, the $ 2 \to 3$ process of
$q g \to q' T \bar{b}$, the $s$-channel process of $q \bar{q}' \to T
\bar{b}$, and the gluon fusion process of $gg\to T\bar{t}$. We found
that the $2 \to 3$ process can be quite important, as its cross
section is about 30\% of the $2\to 2$ one and it is dominant in high
$p_T$ distributions. The $s$-channel and the gluon fusion processes
also show distinctive features in spite of their suppressed cross
sections. In the gluon fusion process of the simplest little Higgs
model, for example, the pseudo-scalar contribution is rather
dominant over the Higgs contribution for relatively light $M_T$.
\end{abstract}
\maketitle

\section{Single Top production}

Recently, new models, dubbed the ``little Higgs'' models, have drawn
a lot of interests as   new candidates for the solution
of the gauge hierarchy problem\,\cite{LH}.
The original idea dates back to 1970's: The lightness of
the Higgs boson is attributed to its being a pseudo Nambu-Goldstone boson,
which is generated when a global symmetry is spontaneously broken\,
\cite{Georgi:yw}.
Unfortunately, the model was not phenomenologically viable
due to the same quadratic divergence of the radiative correction
to the Higgs mass as in the standard model (SM).
Re-armed by the collective symmetry breaking idea,
little Higgs models could explain the little hierarchy problem
since the radiative generation of Higgs
mass occurs at two-loop level.
The one-loop level quadratic divergences
from the SM gauge boson and top-quark loops are
cancelled by those from new heavy gauge boson and heavy $T$-quark loops,
respectively.
Note that the cancellation at one-loop level happens between the
SM particles and the new particles with the same statistics
in the little Higgs model, unlike in supersymmetric theories.
The exactly opposite coupling strength leads to the cancellation.
This one-loop level cancellation is called the little Higgs mechanism.

Although there are many specific little Higgs models according to
the global symmetry breaking pattern\,\cite{various}, we can
classify them into two categories\,\cite{smoking}. The first one is
called the ``product group'' models, in which the diagonal breaking
of two (or more) gauge groups leads to the SM gauge group. The most
representative one is the littlest Higgs model. The second one is
the ``simple group" models, where a single gauge group is broken
into the SM $SU(2)_L$ gauge group. The $SU(3)$ simple group model,
so called the simplest little Higgs model, is one of the most
studied simple group models. In literatures, most of
phenomenological and experimental studies on little Higgs models at
the Large Hadron Collider (LHC) have been focused on the littlest
Higgs model\, \cite{phenomenology,LstATLAS,PPP,perestein:review}.

A new colored heavy $T$-quark is one of the most crucial
ingredients of little Higgs models.
The $T$-quark was introduced to cancel the SM top-quark
contribution to the Higgs boson mass,
which is the largest due to its large Yukawa coupling.
Moreover, this $T$-quark decay pattern shows one novel feature of little Higgs
models --
the neutral decays of $T \to Zt$ and $T \to Ht$
have equal branching ratios\,\cite{Azuelos:2004dm}.
The first experimental signature of little Higgs models at high energy colliders
would be the production of new heavy particles.
However, the $T$-quark is considered to be quite heavy
because of the strong constraint from the electroweak precision data (EWPD):
The $T$-quark should be heavier than about one TeV,
which suppresses its pair production at LHC.
Naively estimated reach
in a single $T$-quark production channel is up to $\sqrt{\hat s}$, 
while the reach in pair production is at most $\sqrt{\hat s}/2$.
Previous studies on the single $T$-quark production were, however, only
on the total cross section of $qb \to q' T$ mediated by the SM
$W$ gauge boson \cite{PPP}.
Even though this $2 \to 2$ process has dominant total cross section,
its $p_T$ distribution is prone to low $p_T$ region, compared to $2\to 3$ process of
$q g \to q' T \bar{b}$.
Also, the $s$-channel process of $q \bar{q}' \to T\bar{b}$
can be quite sizable if the resonant decay of the heavy charged gauge boson
is allowed.

Another process of interest is the gluon fusion with neutral scalar exchanges, which
gives rise to heavy $T$-quark production associated with the SM top quark.
It could be significant if the $T$-quark mass is light enough.
As shall be discussed,
the $T$-quark mass in the simplest little Higgs model
can be considerably light at large $\tbt$,
where $\tbt$ is the ratio of two vacuum condensates.
In addition to the SM Higgs boson, the relatively light pseudo-scalar $\eta$
in the simplest  little Higgs model can also mediate the gluon fusion process.
We shall show that the contribution from $\eta$ exchange can dominate at
small $M_T$.

Various kinematic distributions are also of great interest since the
data-selection at LHC environment prefers high transverse momentum.
We expect that the subdominant processes of the $2\to 3$, the
$s$-channel process, and the gluon fusion will become dominant in
high $p_T$ region. Therefore, the main goal of the paper is to study
comprehensively the single heavy $T$-quark production at LHC,
including kinematic distributions, for both representative models,
the littlest Higgs model and the $SU(3)$ simplest little  Higgs
model.

This paper is organized as follows. In Section \ref{sec:review}, we
review the basic formalism of the littlest Higgs model and the
$SU(3)$ simplest  little Higgs model. We explore the parameter space
where the new heavy $T$-quark can be relatively light, and show that
the $T$-quark mass in the simplest  little Higgs model can be as low
as about 500--700 GeV. Section \ref{sec:Tb} deals with the $T$-quark
production accompanied by light quarks (including $b$ quark) without
resort to the effective-$W$ approximation. We present the formula of
the parton-level cross section for each process. In Section
\ref{sec:ggfusion}, we consider, focusing on the simplest  little
Higgs model, the gluon fusion process of $T t$ production mediated
by the SM Higgs boson and the light pseudo-scalar boson $\eta$.
Section \ref{sec:numeric} presents the numerical results of the
total cross sections and various distributions. We summarize in
Section \ref{sec:conclusions}.

\section{Review of the littlest Higgs model and
the simplest little Higgs model}
\label{sec:review}
\subsection{The littlest Higgs model}
The littlest Higgs model is embedded into a non-linear $\sigma$
model in the coset space of $SU(5)/SO(5)$ with additional local
gauge symmetry $[SU(2) \otimes \thin U(1)]^2$~\cite{Lst,Lst:Han}.
The covariant two-derivative term for the sigma field $\Sigma$ has
the leading term of \beq \label{2dim} \mathcal{L}  _{\Sigma} =
\frac{1}{2} \frac{f^2}{4}
    {\rm Tr} | \mathcal{D}_{\mu} \Sigma |^2 \,,
\eeq
where the covariant derivative is
\beq
\label{covariantD} \mathcal{D}_\mu \Sigma=
\partial_\mu\Sigma + i \sum_{j=1}^2\big[ g_jW_j^a( Q_j^a \Sigma +
\Sigma Q_j^{aT}) + g'_j B_j(Y_j\Sigma + \Sigma Y_j^T) \big].
\eeq
At the scale $\Lambda_S \sim 4 \pi  f$, a symmetric tensor of the
$SU(5)$ global symmetry develops a vacuum expectation
value (VEV) $f\sim 1$ TeV, of which direction is into the $\Sigma_0$:
\begin{equation}
\label{Sigma0} \Sigma_0 = \left( \begin{array}{ccc}
 & & {\mathbf{1}}_{2 \times 2} \\
 &1 & \\
{\mathbf{1}}_{2 \times 2} & &
\end{array}\right).
\end{equation}
The $\Sigma_0$ triggers two kinds of symmetry breaking.
First the global $SU(5)$ symmetry is broken into $SO(5)$,
resulting in 14 massless Nambu-Goldstone
bosons. 
Second, the assumed gauge symmetry
$[SU(2)\otimes \thin U(1)]^2$ is also broken into its diagonal subgroup
$SU(2)_L \otimes U(1)_Y$, identified as the SM gauge group.
The gauge fields $\overrightarrow{W}^{\, ' \mu }$ and $B^{'\mu }$ with
the broken gauge symmetries become massive by eating four Nambu-Goldstone
bosons. 

The non-linear sigma fields are then parameterized by the
Goldstone fluctuations:
\begin{equation}
\Sigma = \Sigma_0 + \frac{2 i}{f} \left( \begin{array}{ccccc}
\phi^{\dagger} & \frac{h^{\dagger}}{\sqrt{2}} &
{\mathbf{0}}_{2\times
2} \\
\frac{h^{*}}{\sqrt{2}} & 0 & \frac{h}{\sqrt{2}} \\
{\mathbf{0}}_{2\times 2} & \frac{h^{T}}{\sqrt{2}} & \phi
\end{array} \right) + {\cal O}\left(\frac{1}{f^2}\right),
\end{equation}
where $h$ is a doublet and $\phi$ is a triplet under the unbroken
$SU(2)_L$.
The gauge fields $\overrightarrow{W}^{\,' \mu}$ and $B^{'\mu}$
are related with the SM gauge fields by
\begin{eqnarray}
\label{eq:Lst:gaugemixing}
    W^\mu = -s \thin W^\mu_1 - c \thin W^\mu_2\,, &\qquad&
    W^{\prime\mu} =  \thin c  \thin W^\mu_1 - s  \thin W^\mu_2\,, \nonumber \\
    B^\mu =\phantom{-} s^{\prime} B^\mu_1 + c^{\prime} B^\mu_2\,, &\qquad&
    B^{\prime\mu} = - \thin c^{\prime} B^\mu_1 + s^{\prime} B^\mu_2\,,
\end{eqnarray}
where $g = g_1 s=g_2  \thin c$ and
$g^{\prime} = g_1^{\prime} s^{\prime}=g_2^{\prime}  \thin c^{\prime}$.

In order to cancel the severe quadratic divergence
from the top quark loop,
the top quark sector is extended with at least one additional
top-quark-like fermion with a heavy mass of the order of $f$.
In a minimal extension,
a pair of $SU(2)_L$-singlet but colored Weyl fermions
$U_L$ and $U_R$ with electric charge $+2/3$ are introduced.
The Yukawa term for
the third generation quarks
and the singlet $u_{3R}$ is \cite{PPP}:
\beq
\label{eq:topyuk:LH}
 {\cal L}_{\rm top} =   - {\lambda_1\over 2}f\, \chi_{L i}^\dagger
  \epsilon_{ijk} \epsilon_{mn} \Sigma_{jm} \Sigma_{kn} u_{3R} - \lambda_2 f\,
   U^\dagger_L U_R + {\rm h.c.}\  ,
\eeq
where  $i,j,k$ run between $1$ and $3$, $m, n=4,5$,
and $ \chi_L^T =\left( -i d_{3L}, i u_{3L}, U_L \right) $\,
\cite{perestein:review}.
The $\Sigma$ field VEV mixes the $u_{L,R}$ fields
with the $U_{L,R}$ fields into
one massive mass-eigenstate $T_0$ (with TeV scale mass) and one
massless eigenstate $t_0$ at this stage:
\bea
  t_{0L} = u_{3L}, &\qquad& t_{0R} =  {\lambda_2 u_{3R}- \lambda_1 U_R\over
                      \sqrt{\lambda_1^2 + \lambda_2^2}},
\no \\
  T_{0L} = U_L, & \qquad& T_{0R} = {\lambda_1 u_{3R} + \lambda_2 U_R\over
                                   \sqrt{\lambda_1^2 + \lambda_2^2}}.
\eea
The electroweak symmetry breaking (EWSB) is triggered by
the Higgs boson VEV $v$, radiatively generated at two-loop level.
This Higgs VEV also
induces small mixing between $t_0$ and $T_0$,
leading to final mass eigenstates $t$ and $T$:
\beq
T = T_0 - x_\lm \frac{m_t}{M_T} t_0,
\quad
t =t_0 +x_\lm \frac{m_t}{M_T} T_0,
\eeq
where $x_\lambda =\lm_1/\lm_2$.
Through this mixing, we have non-zero $W$-$T$-$b$ coupling, though
suppressed by $\sim v/f$.

The phenomenology of the littlest Higgs model at high energy
colliders depends on the following parameters\footnote{Even though
the Higgs triplet VEV $v'$ is also a model parameter, its effect on
the collider phenomenology is negligible due to its smallness from
the condition of positive definite mass squared of the Higgs
triplet.}: \beq \label{eq:param1} f,\quad c,\quad c',\quad x_\lambda,
\eeq where one of $\lm_1$ and $\lm_2$ is traded with the SM
top-quark mass. It is well known that main contributions to the EWPD
are proportional to $c^2$ or $(c'^2-s'^2)$. In the parameter space
around $c \ll 1$ and $c'=1/\sqrt{2}$, new contributions are
therefore suppressed: $f \sim 2$~TeV is allowed in the region of $c
\in [\, 0\, , \, 0.5 \,]$ and $c' \in [\, 0.62\, , \, 0.73\,
]$\,\cite{EWPD,Csaki:variation}.

\subsection{$SU(3)$ simple group model}
\label{subsec:simplest}
Among four popular representatives of
little Higgs models, the $SU(3)$ simplest little Higgs model
has the lowest fine-tuning associated to electroweak symmetry breaking\,
\cite{fine-tuning}.
The theory has an $[SU(3)\times U(1)_X]^2$ global symmetry
and its diagonal subgroup $SU(3) \times U(1)_X$ is gauged.
The kinetic term in the non-linear sigma model is
\begin{equation}
    \mathcal{L}_{\Phi} = \sum_{i=1,2} \left| \left( \partial_{\mu}
    + i g A_{\mu}^a T^a - \frac{i g_x}{3} B_{\mu}^x \right)
    \Phi_i \right|^2,
    \label{eq:SU3LPhi}
\end{equation}
where $\Phi_{1,2}$ are the complex $SU(3)$ triplet scalar fields,
$T^a$ are the $SU(3)$ generators, and $A^a_\mu$ and $B_\mu$ are, respectively, the
$SU(3)$ and $U(1)$ gauge fields. Contrary to the littlest Higgs
model, the two gauge couplings, $g$ and $g_x$, are determined by the SM
gauge couplings: $SU(3)$ gauge coupling $g$ is just the SM $SU(2)_L$
gauge coupling and
\begin{equation}
    g_x = \frac{g'}{\sqrt{1 - t_W^2/3}}\approx 1.05 \, g'.
\end{equation}

When $\Phi_{1}$ and $\Phi_2$ develop VEV's of TeV scale
$\langle \Phi_{1,2} \rangle = f_{1,2}$,
two kinds of symmetry breaking occur.
First, the global symmetry is spontaneously broken into its subgroup of $[SU(2) \times U(1)]^2$,
giving rise to ten Nambu-Goldstone bosons.
Second, the gauge symmetry $SU(3) \times U(1)_X$ is broken into the SM
$SU(2)_L \times U(1)_Y$, as five Nambu-Goldstone bosons are eaten.
Five new gauge bosons with TeV scale masses appear,
a $Z'$ gauge boson (a linear combination of $A^8$ and $B^x$)
and a complex SU(2) doublet $(Y^0,X^-)$.
The surviving Nambu-Goldstone bosons of a complex $SU(2)_L$ doublet $h$ and a SM singlet
$\eta$ are parameterized by
\begin{equation}
    \Theta = \frac{1}{f} \left[
        \left( \begin{array}{cc}
        \begin{array}{cc} 0 & 0 \\ 0 & 0 \end{array}
            & h \\
        h^{\dagger} & 0 \end{array} \right)
        + \frac{\eta}{\sqrt{2}}
        \left( \begin{array}{ccr}
        1 & 0 & 0 \\
        0 & 1 & 0 \\
        0 & 0 & 1 \end{array} \right) \right],
\end{equation}
where $h= (h^0, h^-)^T$ and $f=\sqrt{f_1^2+f_2^2}$.
In the non-linear sigma model, they are related with two scalar fields through
\beq
    \Phi_1 = e^{i t_\beta \Theta} \left( \begin{array}{c}
            0 \\ 0 \\ f c_\beta \end{array} \right)
        , \quad
    \Phi_2 = e^{-i \Theta/t_\beta } \left( \begin{array}{c}
            0 \\ 0 \\ f s_\beta \end{array} \right)
        ,
\eeq
where $\tbt=f_2/f_1$, $\cbt\equiv\cos\beta$ and $\sbt\equiv\sin\beta$.

Due to the $SU(3)$ gauge symmetry, the SM fermions are promoted to a
$SU(3)$ triplet $\chi_L = (u_{L},d_L,iU_L)^T$. For the third generation quarks,
the Yukawa
interaction is
\beq
\label{eq:Yuk:simple} {\cal L}_{\rm top} \,=\, i
\lambda_1 U^\dagger _{R1} \Phi_1^\dagger \chi_L \,+\,i \lambda_2
U^\dagger _{R2} \Phi_2^\dagger \chi_L \,+\, {\rm h.c.},
\eeq
where $U_{R1}$ and $U_{R2}$ are two
additional $SU(2)_L$-singlet quarks,  and
$i$'s guarantee positive mass for fermions. For the first two
generation quarks and all generation leptons, Yukawa couplings
can be imposed differently due to their negligible effects on
radiative corrections to the Higgs mass.
In literatures, two versions
for the fermion embedding are discussed, ``universal" model\,\cite{universal} and
``anomaly-free" model\,\cite{kong}. In the universal embedding, all
three generations have identical quantum numbers. We have heavy
up-type quarks and heavy neutrinos. In the ``anomaly-free"
embedding, anomaly-cancellation is required for easier UV
completion; the quarks of the third generation and three generations
of leptons are put into {$\mathbf{3}$} representations of $SU(3)$,
while the first two generation quarks are put into {$\mathbf{\bar
3}$} of $SU(3)$. We have heavy down-type quarks for the first
two generations. Irrespective of the fermion embedding, these two
heavy quarks have TeV scale masses of\,\cite{smoking}
\beq
\label{eq:MQ} M_Q =
s_\beta \lm_Q f,
\eeq
where $\lm_Q$ is the Yukawa coupling with
$\Phi_2$.

Substituting the VEV's of $\Phi_i$ into Eq.~(\ref{eq:Yuk:simple}),
a linear combination of $U_{R1}$ and $U_{R2}$ is coupled
to $U_L$, which generates a TeV-scale Dirac mass.
Its orthogonal combination has no mass term at this stage,
to be identified with the SM top quark.
The mass eigenstates before the EWSB are then
\bea
t_{0L} = u_{3L}, &\qquad& t_{0R}
       =  {-\lambda_2 \sbt u_{3R} +\lambda_1 \cbt U_R\over
                   \sqrt{\lambda_1^2 \cbt^2+ \lambda_2^2 \sbt^2}},
\no \\
T_{0L} = U_L, & \qquad& T_{0R}
       = {\lambda_1 \cbt u_{3R} + \lambda_2\sbt U_R\over
                   \sqrt{\lambda_1^2 \cbt^2+ \lambda_2^2 \sbt^2}}.
\eea
Finally, the EWSB causes a small mixing between $t_0$ and $T_0$:
\beq
T = T_0 \mp \frac{1}{\sqrt{2}\tbt}\frac{v}{f} t_0 \equiv T_0 - \delta_t t_0,
\quad
t = t_0 \pm \frac{1}{\sqrt{2}\tbt}\frac{v}{f} T_0,
\eeq
where the upper sign is for the anomaly-free model
and the lower sign is for the universal model.

Although different heavy fermion spectrum itself is very exciting,
the single heavy $T$-quark production at LHC is not directly affected
by other heavy fermions.
Their only effect is the parameter space allowed
by the EWPD in each model~\cite{simplest,su3:han:skiba}.
According to Ref.~\cite{simplest},
the universal model is strongly constrained by its
new contributions to
atomic parity violation, leading to $f \gsim 3.9~{\rm GeV}$
at 95\% C.L.,
while the anomaly-free model is more sensitive to LEP II data, leading to $f\gsim 2$ TeV.
In addition, the contributions to EWPD are suppressed by large $\tbt$.
Thus, a large $\tbt$ leads to interesting implications for
heavy fermion spectra.
As shall be shown below, heavy $T$-quark mass decreases with increasing $\tbt$
while the first two generation heavy quark masses
increase with $\tbt$ as in Eq.~(\ref{eq:MQ}).
In what follows, therefore, we examine the parameter space of $f \geq 2$ TeV
and large $\tbt$
for the simplest little Higgs model.

\subsection{Heavy masses and relevant couplings}
In little Higgs models, the SM gauge boson masses are modified with
corrections of the order of $v^2/f^2$. Their effects at
LHC are, however, negligible if we only consider the parameter space
allowed by EWPD. Masses of heavy particles in the littlest and simplest
little Higgs models are summarized in Table~\ref{table:heavymass}.
\begin{table}
\renewcommand{\arraystretch}{1.3}
\caption{\label{table:heavymass}
Masse of heavy particles in the
littlest and simplest little Higgs model. Here $x_\lm=\lm_1/\lm_2$
and $\tbt = f_2/f_1$.}
\begin{ruledtabular}
\begin{tabular}{lll}
 &littlest & simplest\\
\hline
charged heavy gauge boson  & $M_{W_H} = \dfrac{g f}{2 s c}$
& $M_{X^\pm}=\dfrac{g f}{\sqrt{2}}$\\
neutral heavy gauge boson  & $M_{Z_H} = \dfrac{g f}{2 s c}$
& $M_{Y^0}=\dfrac{g f}{\sqrt{2}}$\\
& $M_{B_H} = \dfrac{g' f}{2 \sqrt{5}s' c'}$
& $M_{Z'}=\sqrt{\dfrac{2}{1-t_W^2}}\,g\, f$\\
heavy $T$-quark & $M_T=\left(x_\lm+\dfrac{1}{x_\lm}\right)\dfrac{m_t}{v}f$ &
$M_T = \sqrt{2} \dfrac{\tbt^2+x_\lm^2}{(1+\tbt^2)x_\lm} \dfrac{m_t}{v}\,f$\\
\end{tabular}
\end{ruledtabular}
\end{table}

The dominant single $T$-quark production at LHC is accompanied by a
light quark. The production of $T \bar{b}$ is mediated by the
charged gauge boson $W$ and $W_H$ ($X^\pm$) in the littlest
(simplest) Higgs model. We denote the relevant couplings as \beq \Lg
= - \sum_{V=W,W'} g^V_{ud}\, \bar{d} \gm^\mu P_L u V_\mu -
\sum_{V=W,W'} g^V_{Tb} \bar{T} \gm^\mu P_L b V_\mu \;\;\; + \;\;
{\rm h.c.}\,, \eeq where $W'=W_H$ in the littlest Higgs model and
$W'=X^\pm$ in the simplest  little Higgs model. Here the
Cabibbo-Kobayashi-Maskawa (CKM) matrix factors are omitted. In Table
\ref{table:CCcoupling}, we summarize the relevant couplings for the
charged-currents with the $W$ and $W'$ gauge bosons.

\begin{table}
\renewcommand{\arraystretch}{1.3}
\caption{\label{table:CCcoupling}
The charged current couplings for $W$ and $W_H$ or $X^\pm$.
In the simplest  little Higgs model, the flavor misalignment is
not assumed for simplicity, and
the upper (lower) sign is for the anomaly-free (universal) embedding.}
\begin{ruledtabular}
\begin{tabular}{lll}\label{table:coupling}
 &littlest & simplest\\
\hline
$g^W_{ud}$  & $\phantom{-}\dfrac{g}{\sqrt{2}} \left( 1-\dfrac{\Dt}{2}c^2(c^2-s^2)\right)$
  & $\phantom{-} \dfrac{g}{\sqrt{2}} \left( 1-\dfrac{\Dt}{4\tbt^2}\right)$\\
$g^{W'}_{ud}$ &  $-\dfrac{g }{\sqrt{2}} \dfrac{c}{s}$ &
$\pm \dfrac{g}{2 \tbt}\dfrac{v}{f}$\\
$g^W_{Tb}$ & $ \phantom{-}x_\lm  \dfrac{g}{\sqrt{2}} \dfrac{m_t}{M_T}$
& $- \dfrac{g}{\sqrt{2}}\dfrac{\tbt}{1+\tbt^2}
\left(x_\lm-\dfrac{1}{x_\lm}\right) \dfrac{m_t}{M_T}$\\
$g^{W'}_{Tb}$ & $- x_\lm  \dfrac{g}{\sqrt{2}} \dfrac{m_t}{M_T}\dfrac{c}{s}$ &
$\phantom{-} \dfrac{g}{\sqrt{2}}$\\
\end{tabular}
\end{ruledtabular}
\end{table}

The mass of the heavy $T$-quark plays an important role in its
production cross section. A relatively light $T$-quark helps reduce
fine-tuning in the Higgs boson mass, as well as increasing the
production rates for the single heavy $T$-quark. In the littlest
Higgs model, the $M_T$ is invariant under $x_\lm \leftrightarrow
1/x_\lm$, so that the minimum of $M_T$ occurs when $x_\lm= 1$:
\beq
M_T^{\rm (min)} = 2 \frac{m_t}{v}f \approx 1.42\,f \quad \hbox{in
the littlest Higgs model}.
\eeq With the allowed value of $f \gsim
1.5$ TeV by the EWPD, the heavy $T$-quark is quite heavy in the
littlest Higgs model.

\begin{figure}[t!]
\begin{center}
    \includegraphics[scale=0.8]{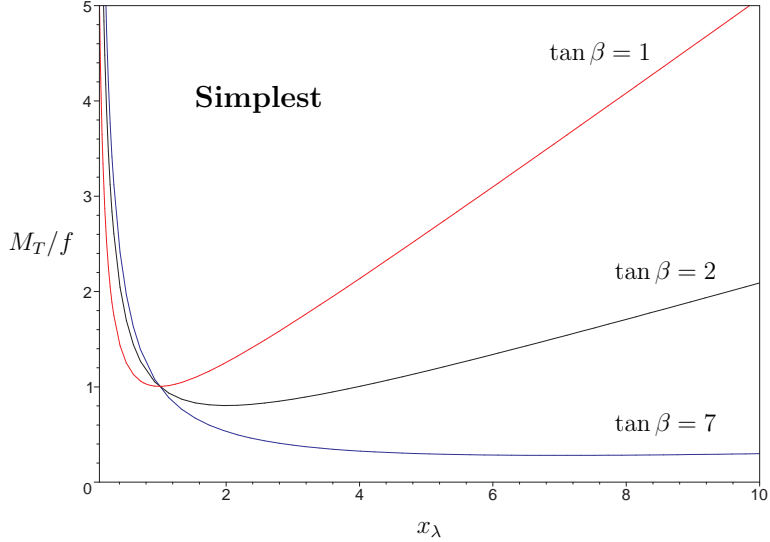}
    \end{center}
    \caption {The heavy $T$-quark mass in unit of $f$ as a function of $x_\lambda$ for
$t_\beta =1,2,7$ in the $SU(3)$ simplest little Higgs model.
    }
    \label{fig:MT:su3}
\end{figure}

In the simplest little Higgs model, $M_T$ can be substantially less.
The minimum of $M_T$ occurs when $x_\lm = \tbt$:
\beq
M_T^{\rm (min)}= 2\sqrt{2}\frac{\tbt}{1+\tbt^2}\frac{m_t}{v}f \quad \hbox{when }
x_\lm=\tbt\,.
\eeq
In Fig.~\ref{fig:MT:su3},
we have plotted $M_T/f$, in the simplest little Higgs model,
as a function of $x_\lm$ for $\tbt=1,2,7$.
It is very interesting that $M_T$ can be significantly reduced for large $\tbt$.
For example, taking $x_\lm=3\,(7)$ and $\tbt=7$ leads to $M_T =0.39\,(0.28)\,f$:
Even for $f=2$ TeV, the heavy $T$-quark mass is only $780\,(563)$ GeV,
which is very accessible at LHC.

\begin{figure}[h!]
\begin{center}
    \includegraphics[scale=0.8]{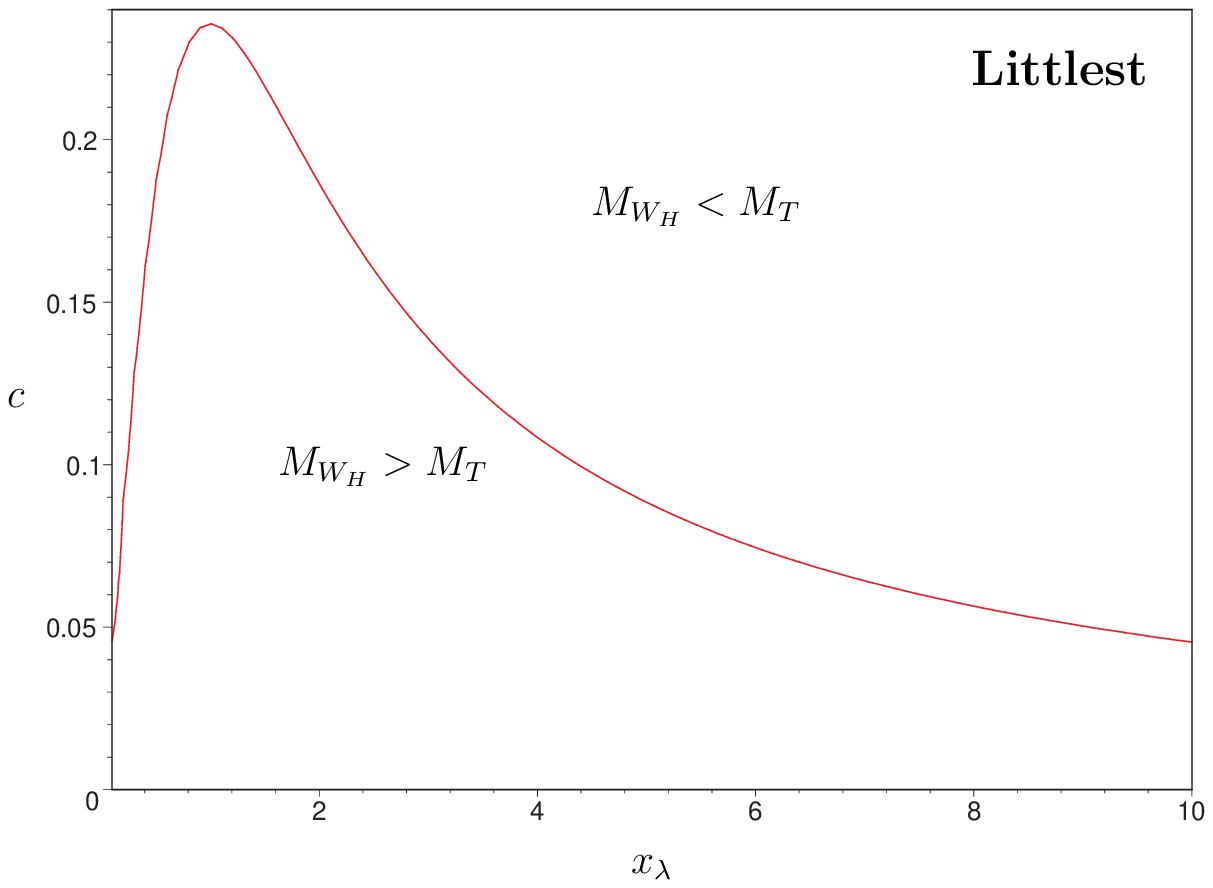}
\includegraphics[scale=0.8]{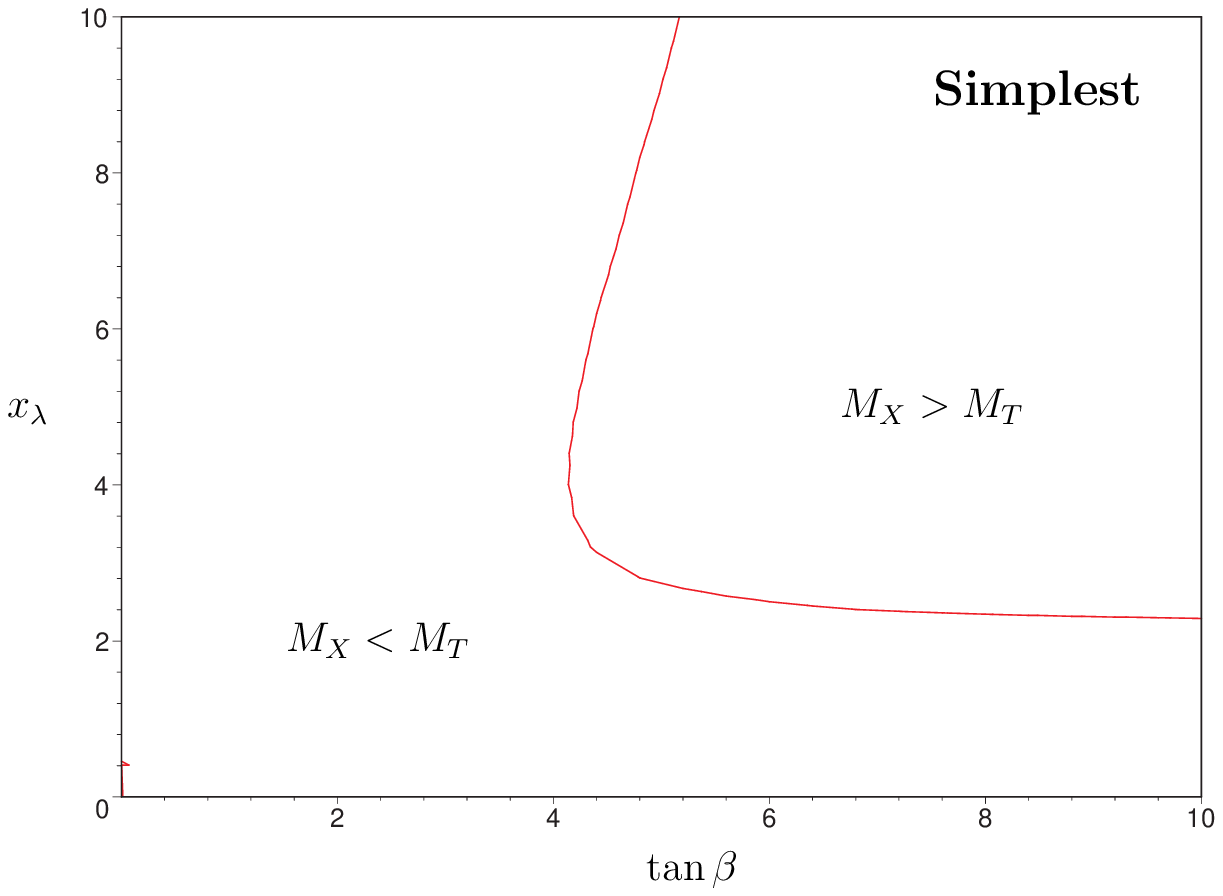}
    \end{center}
    \caption {The parameter space where $W'$ is heavier than the $T$-quark in the
(a) littlest and (b) simplest little Higgs models
    }
    \label{fig:MTMX}
\end{figure}
Another interesting region of parameter space is where the new
charged gauge boson $W'$ becomes heavier than the $T$-quark so that the resonant
production of $u \bar{d} \to W' \to T \bar{b}$ is kinematically
allowed. Figure \ref{fig:MTMX} shows the parameter space where the
$W'$ is heavier than the $T$-quark. In the littlest Higgs model, for example,
the parameter space of $c<0.24$ ($c<0.19$) for $x_\lm=1$ $(x_\lm=2)$
guarantees the $W_H$ heavier than the $T$-quark. Unless $x_\lm$ is
too large, a small value of $c$ ensures the resonant decay of $W_H$
into a $T \bar{b}$ pair. Moreover, a small $c$ is preferred by the
EWPD. Since the minimum of $M_T$ is more or less fixed in the
littlest Higgs model, the condition of $M_T < M_{W_H}$
is satisfied by the heaviness of $W_H$ gauge boson, not by the
lightness of the $T$-quark.

In the simplest little  Higgs model,
the mass of charged gauge boson $X^\pm$ is fixed
at $g f/\sqrt{2}$.
The $M_T$ depends on two parameters, $\tbt$ and $x_\lm$.
As shown in Fig.~\ref{fig:MTMX}(b),
there are two regions in the parameter space
for $M_X > M_T$:
One is where $x_\lm \geq 2 m_t/(g v) \approx 2.18$ and
$\tbt \gsim 4$;
the other is where both parameters are very small (note that
it is barely visible from the figure near the origin).
Considering the EWPD preference for larger $\tbt$ as well as
avoiding too small parameters,
we take $\tbt>1$.
Consequently the resonant decay of $X^\pm$ requires large $\tbt$.
In summary, as $\tbt$ increases
the heavy $T$-quark becomes lighter,
the EWPD constraints become milder,
and the couplings of $X$-$u$-$d$ and $W$-$T$-$b$ become smaller,
while the $X$-$T$-$b$ coupling remains intact.

\section{The single heavy $T$-quark production}
\label{sec:Tb}
\begin{figure}[t!]
\begin{center}
    \includegraphics[scale=1.]{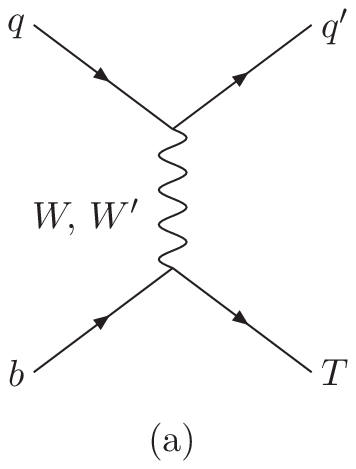}\\
    \includegraphics[scale=1.]{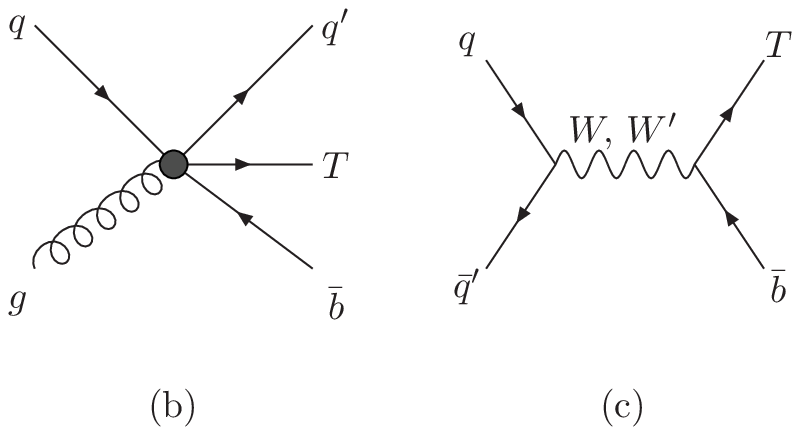}
    \end{center}
    \caption {Feynman diagrams for the heavy $T$-quark production
              at hadron colliders. }
    \label{fig:feyn}
\end{figure}
A heavy $T$-quark can be produced through Feynman diagrams
in Fig.~\ref{fig:feyn}. We denote  the cross section for
diagram (a) by $\sg_{2\to 2}$,
for diagram (b) by $\sg_{2\to 3}$,
and for diagram (c) by $\sg_{\rm s-ch}$.
In general, the total cross sections from
these channels show the following hierarchy:
\beq
\sg_{2\to 2} > \sg_{2\to 3} > \sg_{\rm s-ch}\,.
\eeq
A possible collinear divergence in the $\sg_{2\to 3}$~\cite{Stelzer:1997ns}
is avoided by applying kinematic cuts on the outgoing light quark.
In literatures, only the $\sg(W b \to T)$ is discussed \cite{PPP}.
However, the $t$-channel process of $\sg_{2\to 2} $ generates, in general,
a low $p_T$ distribution.
In high $p_T$ regime, the
$2 \to 3 $ process is expected to become important.

\subsection{Two-body production through $t$ channel diagram}
We study the heavy $T$-quark production associated with a single light quark
through the following parton level processes:
\beq
\label{eq:2-2}
u(p)+ b(p_b) \to d(p') + T(p_T), \quad \mathrm{or}\quad
\bar{d}(p)+ b(p_b) \to \bar{u}(p')+ T(p_T)
\,.
\eeq
For each process, there are two Feynman diagrams
mediated by the $W$ and heavy $W'$ gauge bosons as in Fig.~\ref{fig:feyn}(a).

We define the Mandelstam variables as
\beq
\label{eq:Mandelstam}
\shat = (p + p_b)^2, \quad
\that = (p - p')^2 , \quad
\uhat = (p_b-p')^2 .
\eeq
The spin-averaged amplitude-squared at the parton-level are
\bea
\label{eq:M2ubTd}
\overline{|M|^2}( u b \to T d) &=& |Q_{\that} |^2
\frac{\shat(\shat-M_T^2)}{(\that-\mw^2)^2}\,, \\
\label{eq:M2dbarbTubar}
\overline{|M|^2}( \bar{d} b \to T \bar{u}) &=& |Q_{\that} |^2
\frac{\uhat(\uhat-M_T^2)}{(\that-\mw^2)^2}\,,
\eea
where
\beq
\label{eq:Qt}
Q_{\that} = g^W_{ud} g^{W}_{Tb}
+
g^{W'}_{ud} g^{W'}_{Tb}\frac{\that-\mw^2}{\that-M_{W'}^2}
\,.
\eeq
The couplings $g^{W,W'}_{q q'}$ are listed in Table \ref{table:coupling}.
The parton-level cross section is well known as
\beq
\frac{d \hat\sg}{d \cos\Theta}
=\frac{1}{32 \pi \shat}
\left(
1-\frac{M_T^2}{\shat}
\right)\overline{|M|^2}
\,,
\eeq
where $\Theta$ is the scattering angle of the heavy $T$-quark
in the c.m. frame of partons.
Note that all of the processes have color factors equal to one.

\subsection{Two-body process via $s$-channel diagram and $W'$ decay rate}
For the single heavy $T$-quark production, there is also $s$-channel process
of
\beq
 u(p_1) + \bar{d}(p_2) \to T  (p_T)+\bar{b}(p_{\bar{b}})
\,.
\eeq
Due to the suppressed luminosity of the sea quark,
the cross section
is in general subdominant.
Since this process is
mediated by the heavy $W'$ gauge boson as well as the $W$ gauge boson,
however, a possible resonant mediation by the $W'$ gauge boson
can enhance its contribution.

The spin- and color-averaged amplitude squared is
\beq
\label{eq:M2dbarb}
\overline{|M|^2}( u \bar{d} \to \bar{b} T)
= |Q_{\shat}|^2 \frac{\that(\that-M_T^2)}{(\shat-\mw)^2}\,,
\eeq
where $\shat=(p_1+p_2)^2$, $\that=(p_1-p_{\bar{b}})^2$,
and
\beq
Q_\shat =g^W_{ud} g^{W}_{Tb}
+
g^{W'}_{ud} g^{W'}_{Tb} \frac{\shat-\mw^2}{\shat-M_{W'}^2+ i M_{W'} \Gm_{W'}}\,.
\eeq

Some discussions on the total decay rate of the heavy $W'$ gauge
boson are in order here. In the littlest Higgs model, $W_H$ can
decay into SM fermion pairs or $W h$. In addition, if kinematically
allowed, the $W_H$ can decay into $T \bar{b}$ as well. The total
decay rate is
\beq
\Gm^{\rm Lst} (W_H) =\sum_f \Gamma(W_H \to f \bar{f}')
+\Gamma (W_H \to W h) +\Gamma (W_H \to T \bar{b}) \theta(M_{W'} -
M_T-m_b) \,.
\eeq
Partial decay rates are
\bea
\sum_f \Gamma(W_H \to
f \bar{f}') &=& \frac{g^2 \cot^2\theta}{4\pi}  M_{W_H}, \\ \no \Gamma (W_H
\to W h)&=& \frac{g^2 \cot^2 2\theta}{192 \pi } M_{W_H},
\\ \no
\Gm(W_H \to T \bar{b})
&=&\frac{g^2 x_\lm^2 \cot^2\theta}{16\pi}
\left( \frac{m_t}{M_T} \right)^2 M_{W_H} \left(
1-\frac{M_T^2}{M_{W_H}^2}
\right)\,,
\eea
where $\theta$ is the mixing angle in the gauge sector,
defined by Eq.~(\ref{eq:Lst:gaugemixing}).

In the simplest little  Higgs model,
the $X^\pm$ gauge boson generically couples the new heavy quark
to a SM quark.
For simplicity,
we assume that the first two generation heavy fermions $Q^{1,2}$
are heavy enough so that
the $X^\pm \to Q^{1,2} \bar f$ decay is kinematically prohibited.
If $M_{X^+} > M_T$,
the main decay mode is $X^+ \to T\bar{b}$
with a partial decay rate of
\beq
\Gm^{SU(3)}(X \to T\bar{b}) = \frac{g^2}{16 \pi }M_X
\left(
1-\frac{M_T^2}{M_X^2}
\right)\,.
\eeq
When $X^+ \to T\bar{b}$ is not kinematically allowed,
the decay into the SM fermions $X^+ \to f\bar{f}'$ becomes dominant.
The decay rate summed over the SM fermions is
\beq
\sum_{f,f'} \Gm (X \to f \bar{f}')
=\frac{g^2}{8 \pi t_\beta^2} \frac{v^2}{f^2} M_X\,,
\eeq
which is suppressed by small $v^2/f^2$ and large $\tan\beta$.
This is possible due to the small mixing between the SM fermion
and the heavy fermion.

\subsection{Three-body production}
\begin{figure}[t!]
\begin{center}
    \includegraphics[scale=1.0]{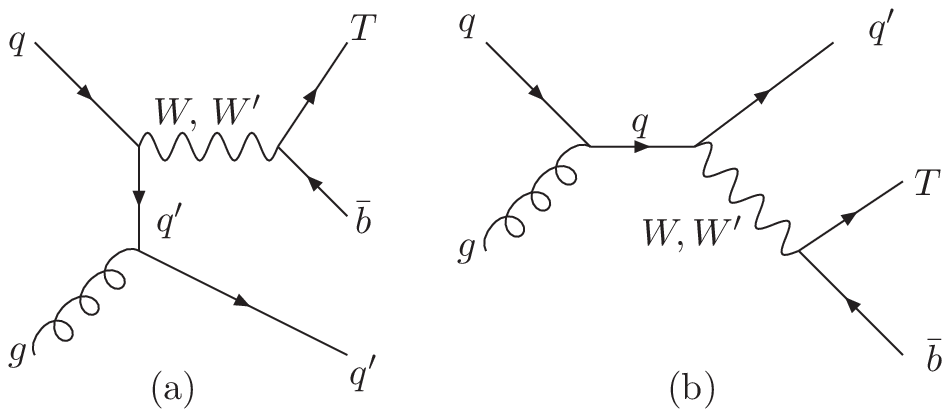} \\[1cm]
    \bigskip
    \includegraphics[scale=1.0]{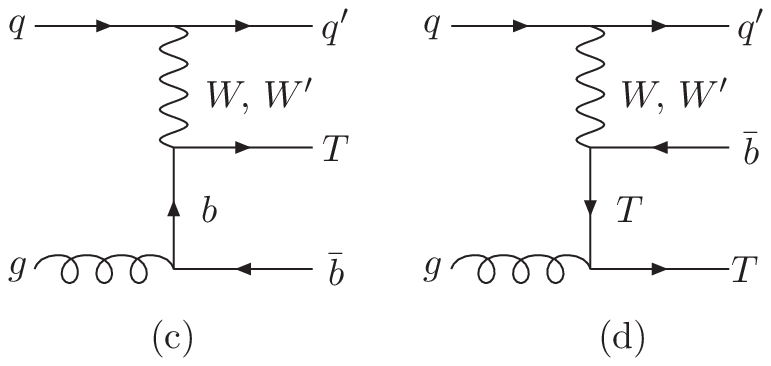}
    \caption {Feynman diagrams for $g u \to d \bar{b} T$.
    }
    \label{fig:Feyn3}
        \end{center}
\end{figure}
The heavy $T$-quark production through $q g \to T \bar{b} q'$
can become dominant in the high $p_T$ region.
Since we are interested in the kinematic distributions,
we compute the differential cross section
without resort to the effective-$W$ approximation.
Instead, we consider the whole set of Feynman diagrams,
mediated by $W$ and $W'$ gauge bosons,
as depicted in Fig.~\ref{fig:Feyn3}.
For the process of
\beq
u(p_1) +  g(p_2) \to d (k_1) +  T(k_2) + \bar b (k_3),
\eeq
the amplitude-squared
(after summing over the final-state colors and helicities, and averaging over
initial-state colors and helicities) is given by Eq.~(\ref{eq:23ampsq})
in the Appendix.
The parton-level cross section is
\beq
d \hat{\sg} = \frac{1}{2^{10} \pi^4}\overline{|M|^2}
\di\eta \, \di\zeta\,\di\alpha\,\di\cos\beta
\,,
\eeq
where the definitions and allowed range of kinematic variables
($\eta$, $\zeta$, $\alpha$, and $\beta$) are also referred to the Appendix.

\section{$ g g\to T \bar{t}$ mediated by scalars}
\label{sec:ggfusion}
\begin{figure}[t!]
\begin{center}
    \includegraphics[scale=1.0]{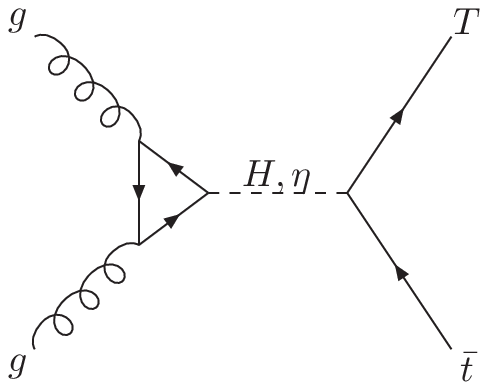}
    \caption {Feynman diagrams for $g g \to T \bar{t}$.
    }
    \label{fig:Feyngg}
        \end{center}
\end{figure}
Due to the high c.m. energy of 14 TeV at LHC,
the gluon luminosity dominates over quark luminosity
unless some very heavy particles are produced.
As discussed before, the simplest little  Higgs model can accommodate
a relatively light $T$-quark with a mass as low as 600 GeV.
In addition, the SM Higgs boson $H$ and the pseudoscalar $\eta$
couple the SM top quark $t$ and the new heavy $T$-quark,
through the mixing between $t$ and $T$ induced by
the EWSB.
Phenomenological impacts and signatures of the $\eta$ boson
at the LHC and other future colliders have been discussed in Ref.~\cite{eta:kilian}.
In this section, we study in the simplest  little Higgs model
the gluon-fusion production of the heavy $T$-quark
associated with $\bar{t}$ quark, as depicted in Fig. \ref{fig:Feyngg}.

We parameterize the Yukawa terms as
\beq
\Lg_Y = - \frac{m_f}{v} y^H_f  \bar{f}  f H
- \frac{m_f}{v} y^{\eta}_{ f} \; i \bar{f} \gm_5 f \eta
+\kp_H (H \overline{T}P_L t + h.c.)
+ \kp_\eta \left(i \eta \overline{T} P_R t + h.c. \right),
\eeq
where $P_{R,L} = (1 \pm \gm_5)/2$, and
\bea
\label{eq:y-coupling}
y^H_t &=& 1 + \mathcal{O}\left( \frac{v^2}{f^2} \right), \quad
y^\eta_t =
\frac{ \big[ \tbt^4+ 2 \tbt^2 (x_\lm^2 -1) -x_\lm^2 \big]}{\sqrt{2} \tbt(\tbt^2+x_\lm^2)}
\frac{v}{f},
\\ \no
y^H_T &=& - \frac{x_\lm^2 (\tbt^2+1)}{2 (\tbt^2+x_\lm^2)^2 } \frac{v^2}{f^2},
\quad
y^\eta_T = \frac{\tbt (x_\lm^2-1)}{\sqrt{2} (\tbt^2+x_\lm^2)} \frac{v}{f}\,,
\\ \no
\kp_H &=& \frac{\tbt}{\tbt^2+1} \frac{ x_\lm^2-1}{ x_\lm } \frac{m_t}{v},
\quad \kp_\eta = \frac{m_t}{v}
\,.
\eea

The couplings of $\phi$-$Q_{1,2}$-$Q_{1,2}$ and $\phi$-$T$-$T$ ($\phi=H,\eta$)
also affect the $g$-$g$-$\phi$ coupling through the
triangle diagram at one-loop level.
Here $Q_{1}$ and $Q_{2}$ are the heavy quarks for the first two generations.
In both universal and anomaly-free embedding cases,
the $H$-$Q_{1,2}$-$Q_{1,2}$ couplings vanish,
while the $\eta$-$Q_{1,2}$-$Q_{1,2}$ couplings show a difference\,\cite{smoking}:
\beq
y^H_{Q_{1,2}} =0, \quad
y^\eta_{Q_{1,2}}=\left\{
                 \begin{array}{ll}
                  \dfrac{1}{\sqrt{2(\tbt^2+x_{\lm_1, \lm_2}^2)}}\dfrac{v}{f},
                      & ~~~ \hbox{for anomaly-free;} \\
             -\dfrac{1}{\sqrt{2}\tbt}\dfrac{v}{f}, & ~~~\hbox{for universal.}
                   \end{array}
                 \right.
\eeq
Here the $x_{\lm_1,\lm_2}$ are, of $\mathcal{O}(1)$, the ratios of two Yukawa couplings
in the first two generations.

We summarize the Feynman rules for a gluon pair ($G^\mu$ and $G^\nu$)
with a scalar boson as \cite{Choi}
\bea
G^\mu(k_1)-G^\nu(k_2)-H &:& \phantom{-} i c_H \dt^{ab}
\left[
k_1 \cdot k_2\eta^{\mu\nu} -k_1^\nu k_2^\mu
\right] , \\ \no
G^\mu(k_1)-G^\nu(k_2)-\eta &:& -i c_\eta \dt^{ab} \epsilon_{\mu\nu\rho\sigma}
k_1^\rho k_2^\sigma ,
\eea
where $a$ and $b$ are the gluon color indices.
The factors $c_H$ and $c_\eta$ are given by
\bea
c_H &=&- \frac{\alpha_s}{4 \pi  v} \sum_f y^H_f \Fs (\tau_f), \\ \no
c_\eta&=& -\frac{\alpha_s}{4 \pi  v} \sum_f y^\eta_f \Fp (\tau_f) \;,
\eea
where the dimensionless loop factors of $F^{H,\eta}_{1/2}$
are~\cite{ggh}
\beq
    F^{H}_{1/2} = -2\tau [1 + (1-\tau)f(\tau)],\quad
    F^{\eta}_{1/2} = -2\tau f(\tau),
\eeq
with
\begin{equation}
    f(\tau) = \left\{ \begin{array}{lr}
        [\sin^{-1}(1/\sqrt{\tau})]^2, & \tau \geq 1 \\
        -\frac{1}{4} [\ln(\eta_+/\eta_-) - i \pi]^2, & \, \tau < 1
        \end{array}  \right.
\end{equation}
and
\begin{equation}
    \tau_f =  \frac{4m_f^2}{m_H^2}, \qquad
    \eta_{\pm} = 1 \pm \sqrt{1-\tau}.
\end{equation}
In the limit of large $\tau$, {\it i.e.}, when the particle in the loop is
much heavier than $H$ or $\eta$, the loop factors approach constant values:
\begin{equation}
\label{Fasympt}
    \Fs \to -\frac{4}{3}, \qquad \Fp \to -2.
\end{equation}

For the process of
\beq
g(k_1, \varepsilon_1^\mu) + g(k_2,\varepsilon_2^\nu) \to T(p_T) + \bar{t}(p_T),
\eeq
the total amplitude is
\beq
M ( gg\to T\bar{t}) = \dt^{ab}\dt^{cd}\overline{u}_T
\left[ -
g_S
\left\{
\frac{\eta_{\mu\nu}}{2}-\frac{k_{1\mu} k_{2\nu}}{\shat} \right\}P_L
+ i g_P \frac{\es_{\mu\nu\rho\sg}k_1^\rho k_2^\sg}{\shat}P_R
\right]v_t \varepsilon_1^\mu \varepsilon_2^\nu
\,,
\eeq
where $a$ and $b$ ($c$ and $d$) are the color indices
of the gluons (of $T$ and $t$) and
\beq
g_S = c_H \kp_H \frac{\shat}{\shat-m_H^2},
\quad
g_P = c_\eta \kp_\eta \frac{\shat}{\shat-m_\eta^2}
\,.
\eeq
The spin- and color-averaged amplitude squared is
\beq
\overline{\left|M \right|^2} (g g \to T\bar{t})
= \frac{3}{64} \big( g_S^2+ g_P^2\big)(\shat-M_T^2-m_t^2)
\,.
\eeq
Finally, the parton level cross section is
\beq
\hat{\sg} = \frac{\lm^{1/2}(1,M_T^2/\shat,m_t^2/\shat)}{16 \pi \shat}
\overline{\left|M \right|^2}
\,.
\eeq

\section{total cross sections and kinematic distributions}
\label{sec:numeric}
As discussed in the Introduction,
the usual evaluation of a single heavy $T$-quark production
by considering only the $q b \to T q'$
can be underestimated.
The cross section of $2 \to 3$ process through
$q g  \to T \bar{b} q'$ is sizable,
and can in fact generate higher $p_T$ distributions.

\begin{figure}[t!]
\begin{center}
    \includegraphics[scale=0.9]{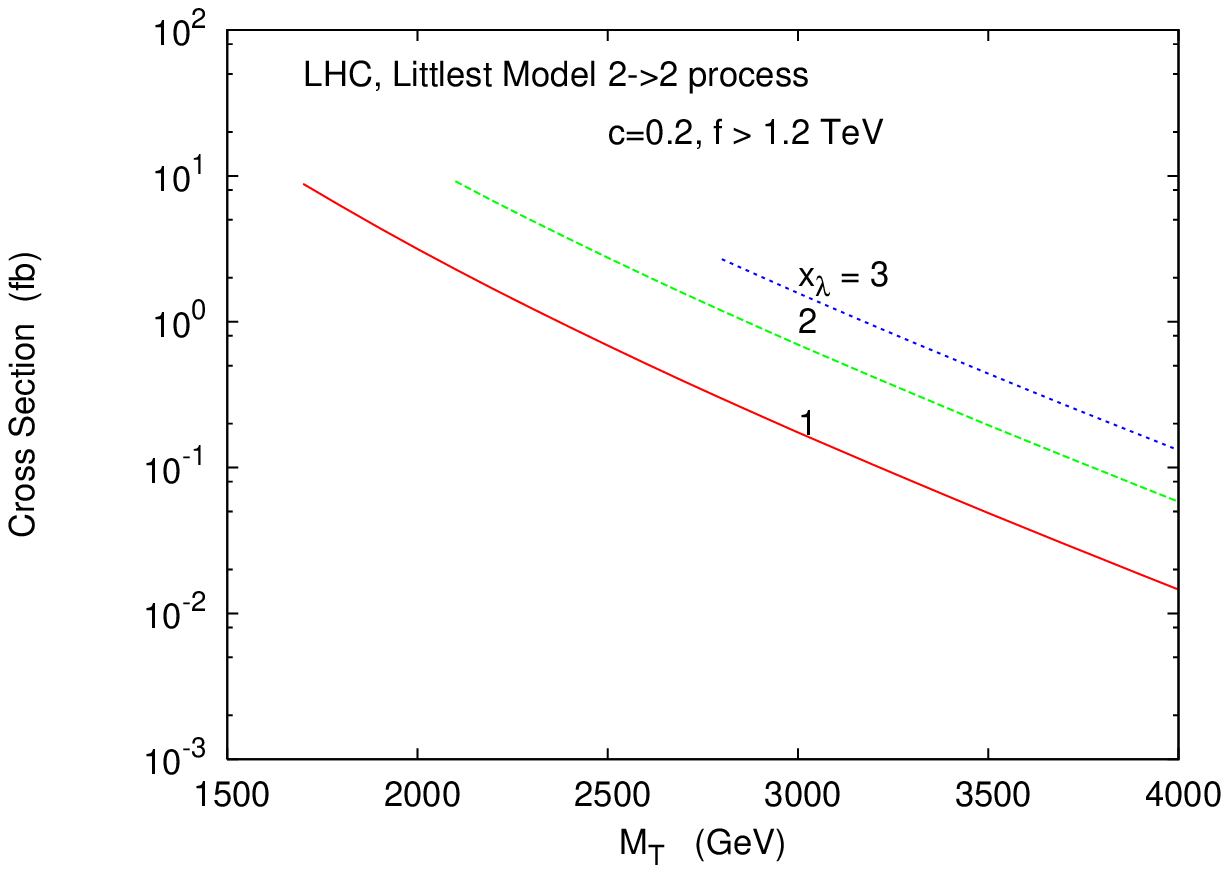}
    \includegraphics[scale=0.9]{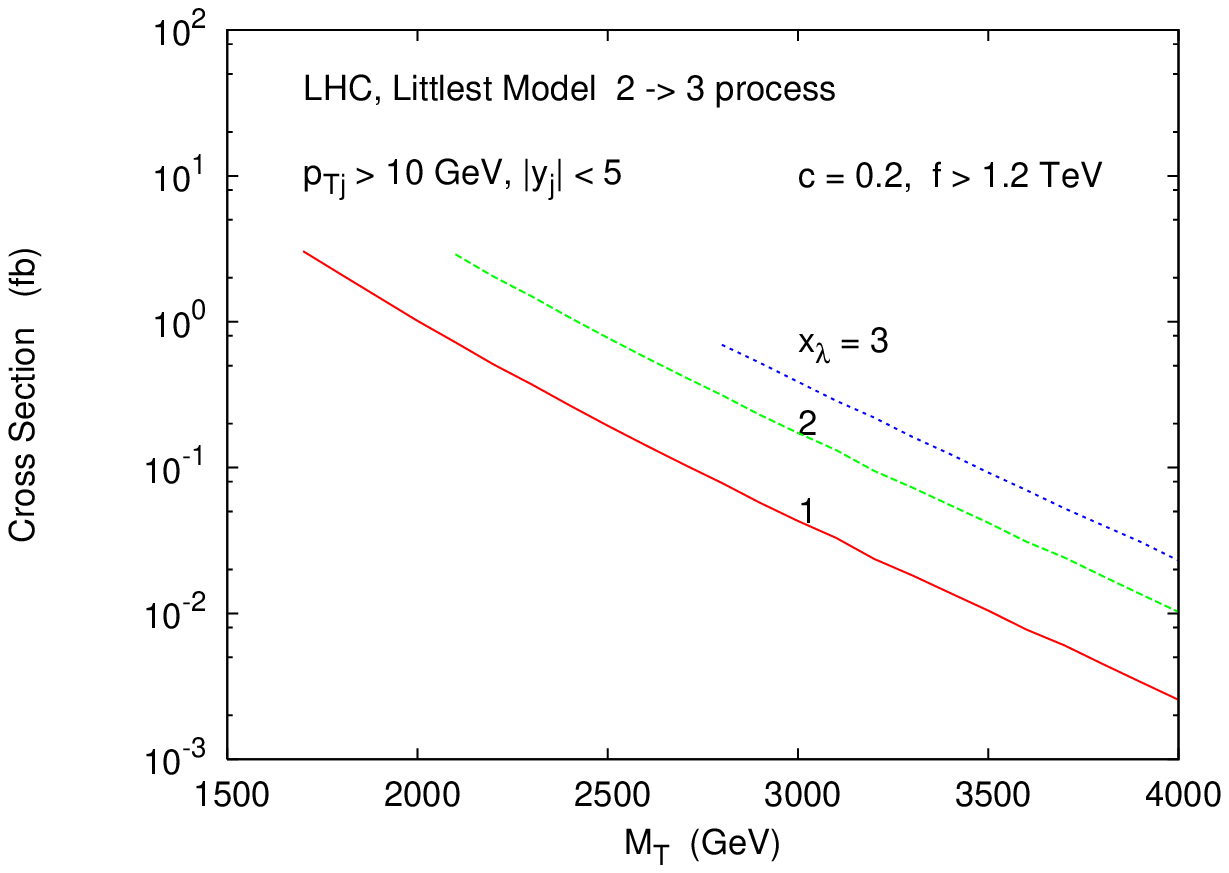}
    \end{center}
    \caption {Production cross section for $T$
as a function of $M_T$ in the littlest Higgs model. We set $c=0.2$
while taking all $f$ above 1.2 TeV. We have taken into account both
$T$ and $\overline{T}$ quarks.
    }
    \label{fig:littlest:totsg}
\end{figure}
In Fig.~\ref{fig:littlest:totsg}, we show, in the littlest Higgs
model, the total cross sections of the single $T$-quark production at LHC
($pp \to Tj(j')X + \overline{T} j (j') X$) as a function of $M_T$
for $x_\lm = 1,2,3$. We have set $c=0.2$, to suppress extra
contributions to EWPD, and taken $f>1.2 $ TeV. Note that each line
starts at different $M_T$ since we have constrained $f$ to be above
1.2 TeV, which is marginally allowed by the EWPD. For the $2 \to 2$
production, we have included all the $t$-channel and $s$-channel
sub-processes, among which the $t$-channel ones dominate. This is
expected since the resonant contribution comes from the narrow
window around $\shat \simeq M_{W'}^2$: The total cross section is not
dramatically enhanced. The total cross section for $2 \to 3$ process
is smaller, but not negligible: $\sg_{2\to 3}$ is about $30 \sim 40\%$ of
$\sg_{2 \to 2}$.

\begin{figure}[t!]
\begin{center}
    \includegraphics[scale=0.9]{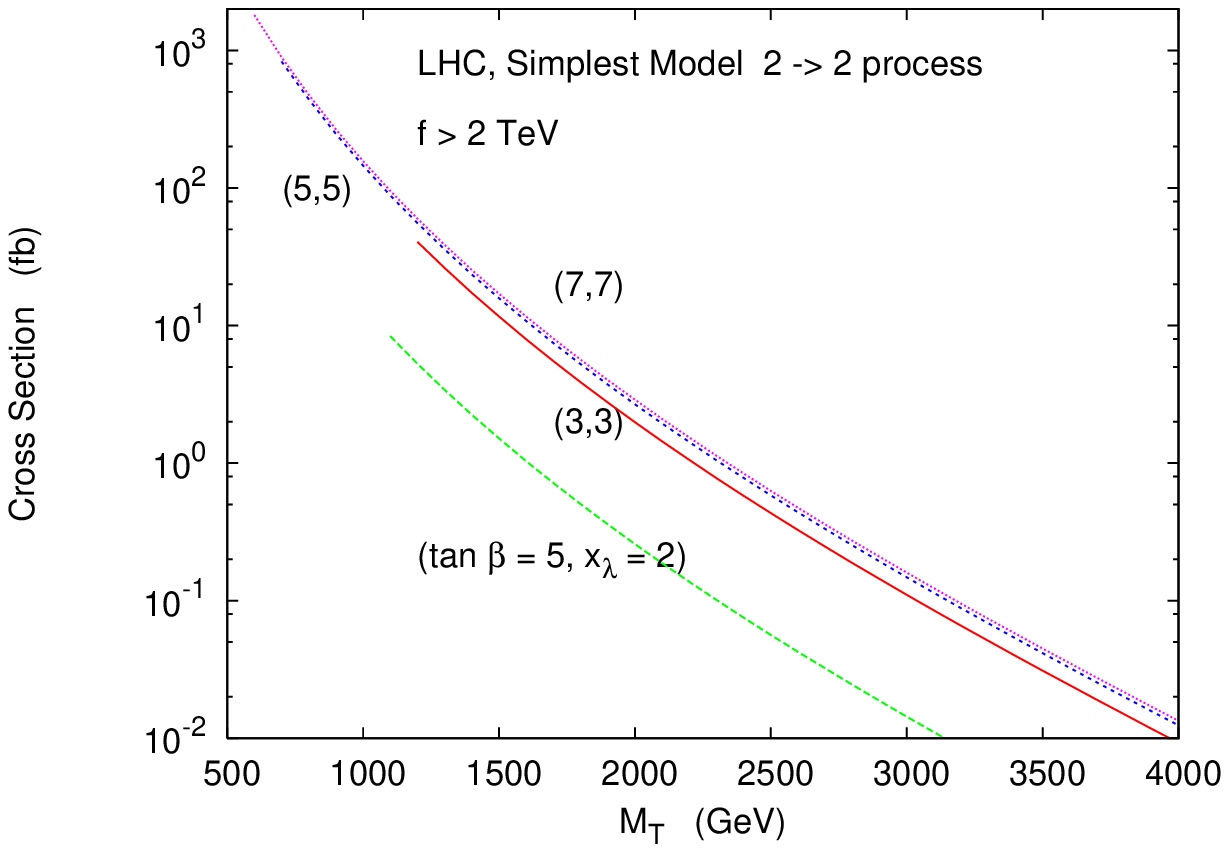}
    \includegraphics[scale=0.9]{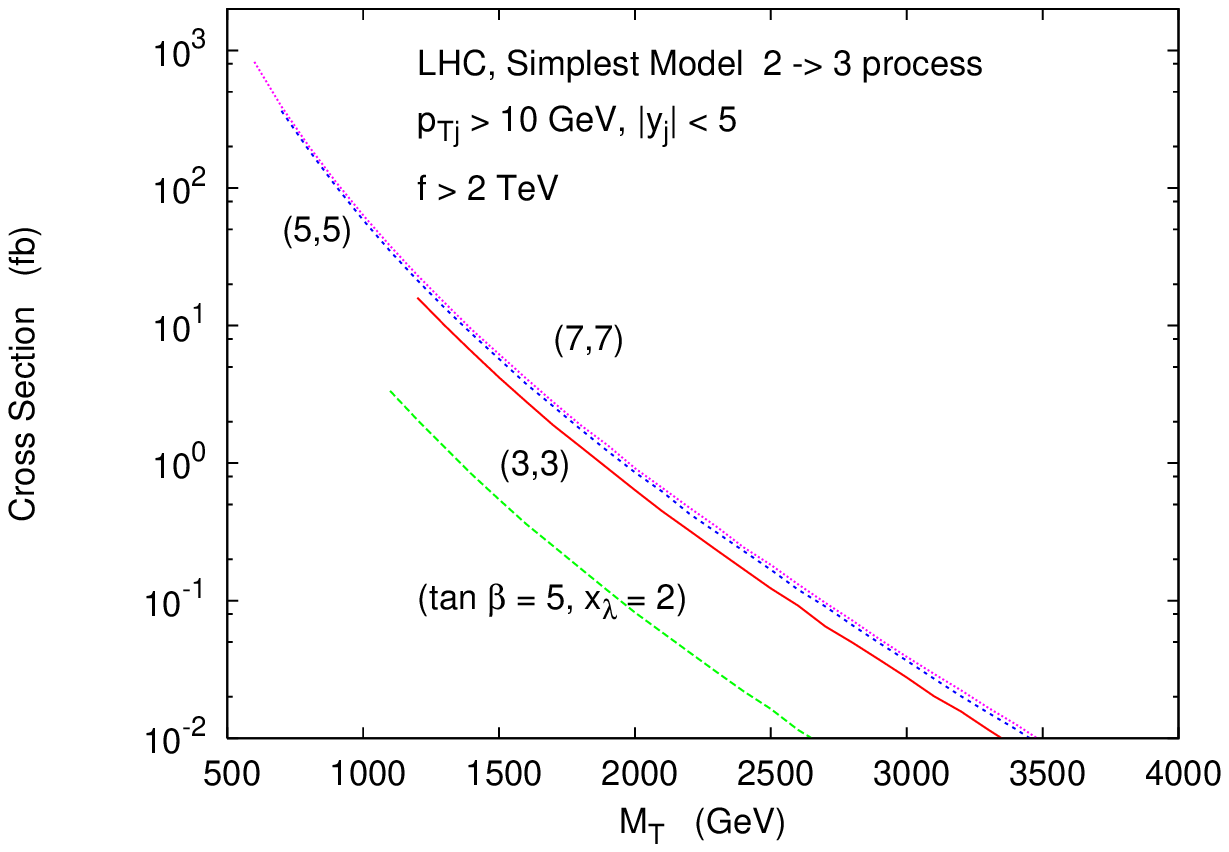}
    \end{center}
    \caption {Production cross section for $T$
as a function of $M_T$ in the simplest  little Higgs model.
We take $f>2$ TeV and $(\tbt,x_\lm)=(5,2),(3,3), (5,5) (7,7)$.
We have taken into account both $T$ and $\overline{T}$ quarks.
    }
    \label{fig:simplest:totsg}
\end{figure}

In Fig.~\ref{fig:simplest:totsg}, we show the same plots for the
simplest  little Higgs model. Here we take a conservative bound for
$f >2$ TeV, consistent with the EWPD. The minimum of $f$ also leads
to the minimum of $M_T$ with the given $\tbt$ and $x_\lm$. The true
minimum of $M_T$ occurs when $\tbt = x_\lm$, which we call the
``$M_T$-minimal case". In addition, the $M_{T}^{\rm (min)}$
decreases with increasing $\tbt$. For example, $M_T^{ \rm (min)}
\simeq 770$ GeV for  $(\tbt,x_\lm)=(5,5)$, while $M_T^{ \rm (min)}
\simeq 560$ GeV for  $(\tbt,x_\lm)=(7,7)$. Another interesting
feature is that once $\tbt=x_\lm$, the total cross section of either
$2 \to 2$ process or $2 \to 3$ process does not change much with the
change of  $\tbt$ value, as explicitly shown in
Fig.~\ref{fig:simplest:totsg}. Compared with the littlest Higgs
model, the $M_T$-minimal simplest little Higgs model has compatible
total cross section with the same $M_T$. In the simplest little
Higgs model, the total cross section for the $2 \to 3$ process is
also about $30\sim 40\%$ of the $2 \to 2$ process.

\begin{figure}[t!]
\begin{center}
    \includegraphics[scale=0.9]{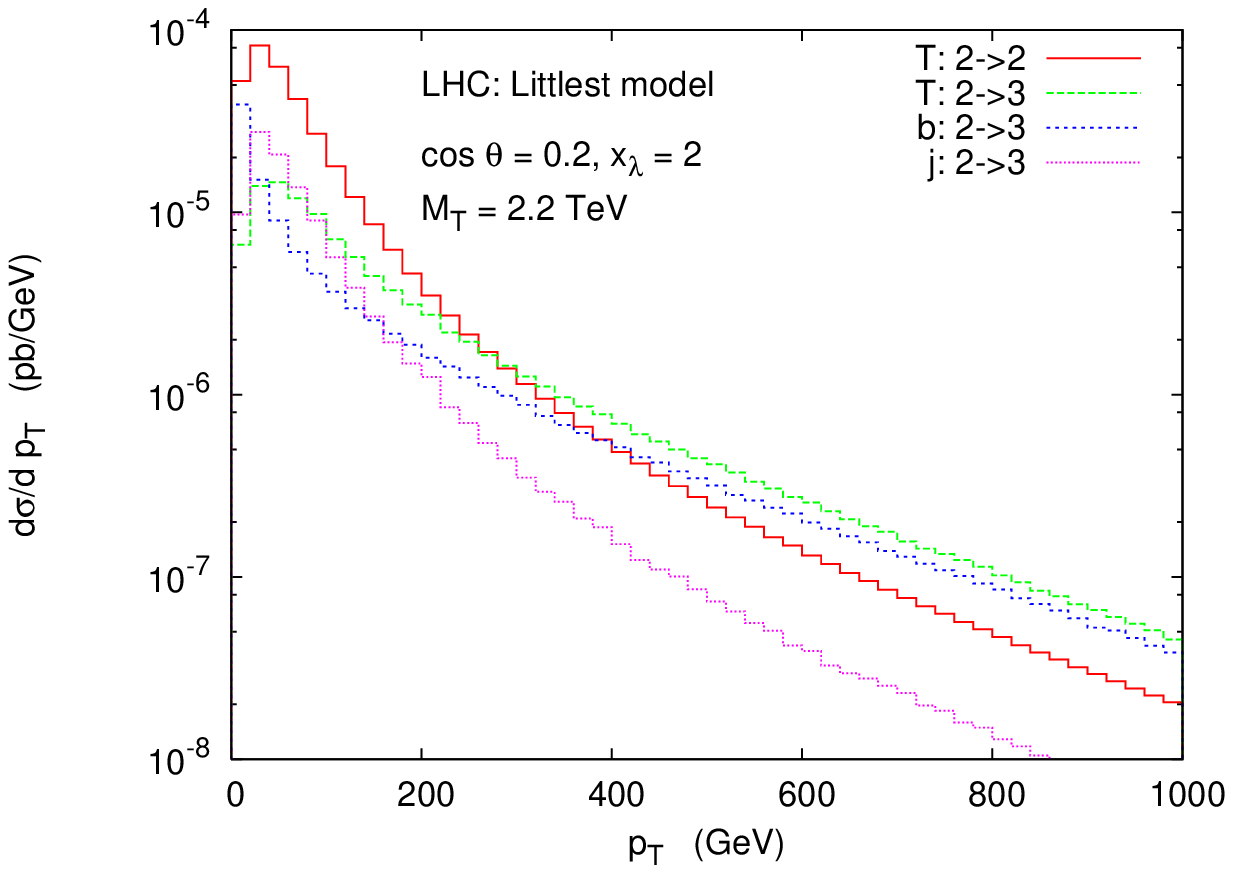}
    \end{center}
    \caption {$p_T$ distributions for the single $T$-quark production
at LHC, in the littlest Higgs models.
    }
    \label{fig:pt:littlest}
\end{figure}
Figure \ref{fig:pt:littlest} exhibits, in the littlest Higgs model,
the $p_T$ distributions for the single $T$-quark production at LHC. The
solid (red) line shows the $p_T$ distribution of the heavy $T$-quark
in the $2 \to 2$ process, while the dashed (green) line shows the
same $p_T$ distribution in the $2 \to 3$ process. In the $2 \to 3$
process, we also show the $p_T$ distributions for $b$ quark
(long-dotted or blue) and for jet (dotted or pink). For the low
$p_T$ data, which constitutes major part of the total cross section,
the $2 \to 2$ process has much larger distributions compared to the
$2 \to 3$ process. For high $p_T$ (above $\sim 300$ GeV) data,
however, the $2 \to 3$ process begins to dominate.

\begin{figure}[t!]
\begin{center}
    \includegraphics[scale=0.9]{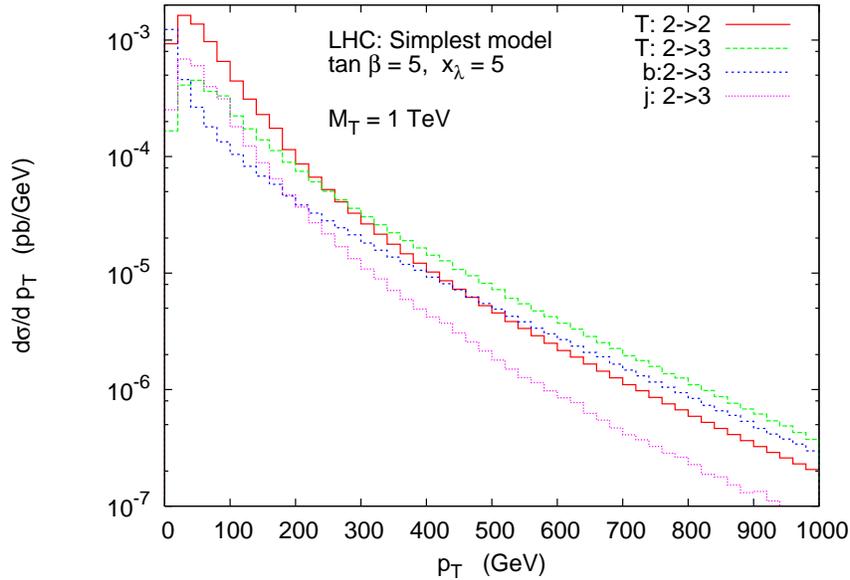}
    \end{center}
    \caption {$p_T$ distributions for the single $T$ production
at LHC, in the simplest little  Higgs models.
    }
    \label{fig:pt:simplest}
\end{figure}

In Fig.~\ref{fig:pt:simplest}, we present the same $p_T$
distributions in the simplest little  Higgs model. Here we fix
$\tbt=x_\lm=5$ and $M_T=1$ TeV. The behavior of each distribution is
similar to the littlest Higgs model case: In the high $p_T$ region,
$2 \to 3$ process becomes more important.

\begin{figure}[t!]
\begin{center}
    \includegraphics[scale=0.9]{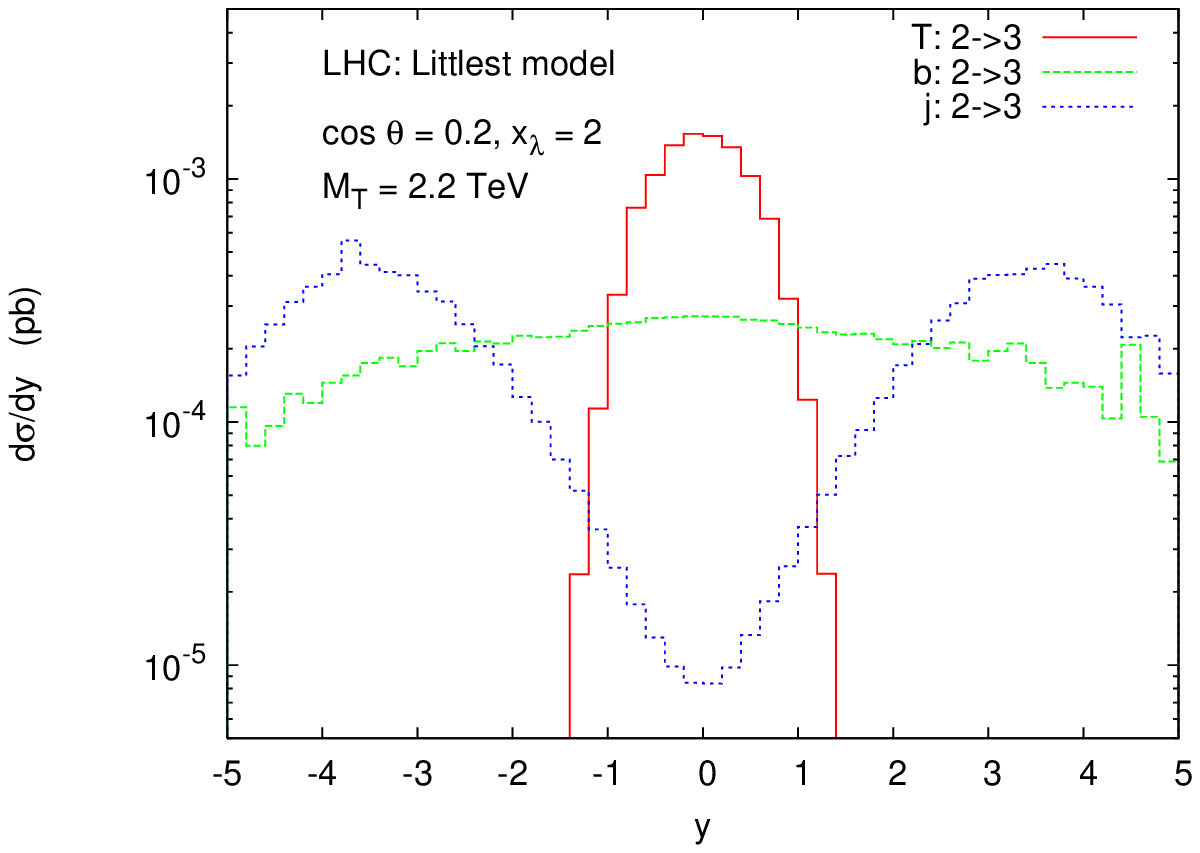}
    \includegraphics[scale=0.9]{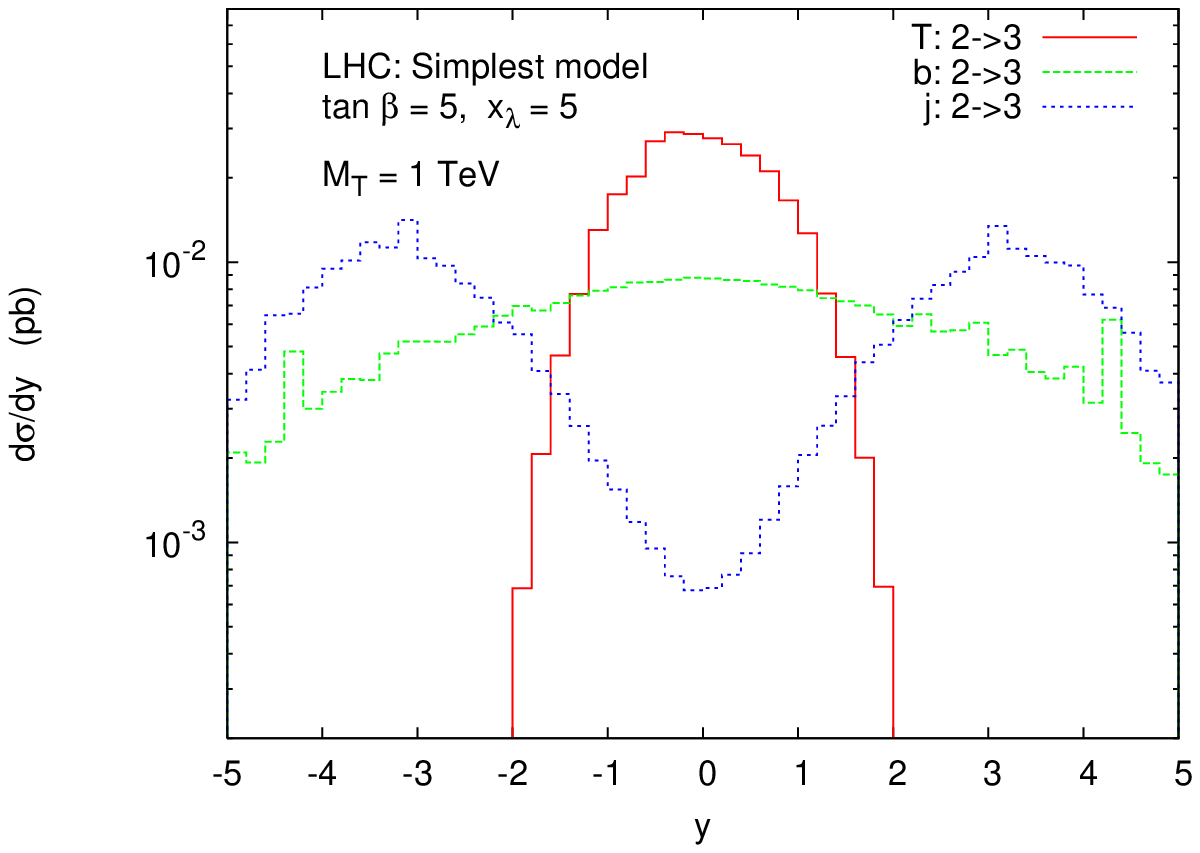}
    \end{center}
    \caption {$p_T$ distributions for the single $T$ production
at LHC, in the littlest and simplest little Higgs models.
    }
    \label{fig:rapidity}
\end{figure}
In Fig.~\ref{fig:rapidity}, we present the rapidity distributions
for out-going particles in the $2 \to 3$ process at LHC, in both the
littlest and simplest little  Higgs models. For the littlest Higgs
model, we set $c=0.2$, $x_\lm=2$, and $M_T =2.2$ TeV, while for the
simplest little Higgs model we set $\tbt =x_\lm=5$ and $M_T =1$ TeV.
We see an interestingly similar tendency: The heavy $T$-quark is
likely produced in the central region, while the $b$ quark
distribution is rather flat. The other light quark jet has the
rapidity distribution of  M-shape.

\begin{figure}[t!]
\begin{center}
    \includegraphics[scale=0.9]{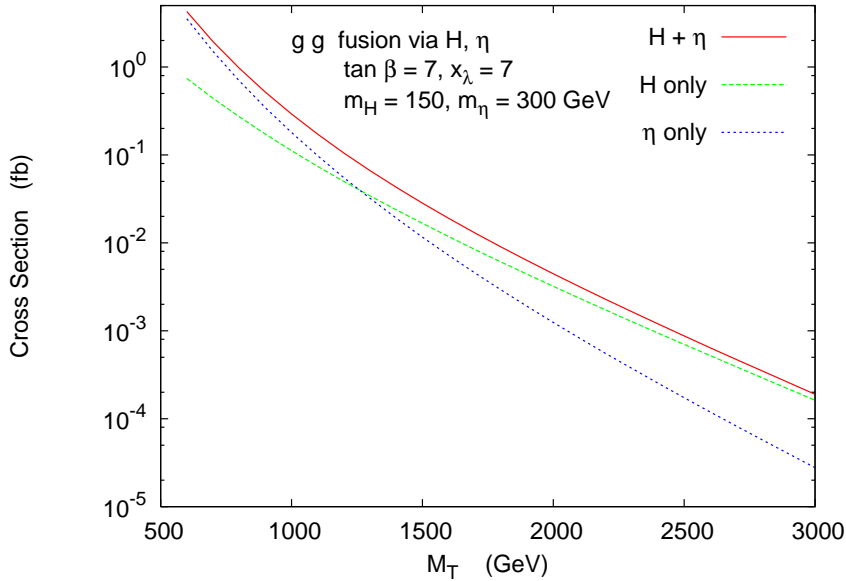}
    \end{center}
    \caption {In the simplest little Higgs model,
    total cross section of the gluon fusion production
of a single $T$-quark accompanied by the SM top quark.
We have taken into account both $T \bar t$ and $\overline{T} t$ channels.
    }
    \label{fig:ggfusion}
\end{figure}
Finally we show,
in Fig.~\ref{fig:ggfusion}, the total cross section of the gluon fusion production
of a single $T$-quark
accompanied by the SM top quark.
We have fixed $\tbt=x_\lm=7$, $m_H=150$ GeV, and $m_\eta =300$ GeV.
The total cross section of the gluon fusion process is $\mathcal{O}(10^{-3})$
of the dominant cross section of the $2\to 2$ process.
In the region of small $M_T$ (or small $f$),
the $\eta$ contribution is dominant over the SM Higgs contribution.
This is expected since the $\eta$-$t(T)$-$\bar{t}(\bar{T})$ couplings of
$y^\eta_t$ ($y^\eta_T$) increase with smaller $f$ and large $\tbt$,
as can be seen in Eq.~(\ref{eq:y-coupling}).
For $M_T \gsim 1.2$ TeV,
the Higgs contribution becomes dominant over the $\eta$ contribution.

\section{Conclusions}
\label{sec:conclusions}

Little Higgs models can solve the little hierarchy problem that
10 TeV cut-off in the SM requires a fine-tuning to the Higgs boson mass of
$\mathcal{O}(100)$ GeV
due to the
quadratic divergence of the radiative Higgs boson mass at one-loop
level. In little Higgs models, the Higgs boson is a pseudo
Nambu-Goldstone boson of a large global symmetry, and collective
symmetry breaking mechanism prohibits the one-loop level radiative
corrections to the Higgs boson mass. Phenomenologically, new heavy
gauge bosons and heavy $T$-quark induce the radiative contributions
which cancel the SM contributions. The heavy $T$-quark, which is
colored, $SU(2)_L$-singlet and vector-like, is one of the most crucial
ingredients, which was introduced to cancel the largest contribution
of the SM top quark to the Higgs boson mass. Its production at LHC
deserves a comprehensive study. In two representative little Higgs
models, $i.e.$, the littlest Higgs model and $SU(3)$ simplest Higgs
model, we have studied the single heavy $T$-quark
 production without any approximation.

We have first searched for the parameter space where
the heavy $T$-quark is relatively light so that
it can be abundantly produced at LHC.
We found that it is possible in the $SU(3)$ simplest little Higgs model.
For example, the $M_T \simeq 560$ GeV is allowed by the EWPD
when $\tbt = x_\lm =7$,
where $\tbt\,\,(x_\lm)$ is the ratio of two vacuum condensates
(Yukawa couplings).
Moreover, the large $\tbt$ is preferred by EWPD.

In literatures, only the $2 \to 2$ process of $q b \to q' T$
has been studied for the single heavy $T$-quark production.
We have extended to the $ 2 \to 3$ process of $q g \to q' T \bar{b}$,
the $s$-channel $q \bar{q}' \to T \bar{b}$, and the gluon fusion process
of $gg\to T\bar{t}$.
The total cross section of $2 \to 3$ process is substantial,
about 30\% of that of $2\to 2$ process.
When we studied various kinematic distributions,
we found in both little Higgs scenarios
that the $2\to 3$ process becomes more important in higher $p_T$ region.
For $p_T \gsim 250$ GeV, the $2\to 3$ process already dominates over the
$2\to 2$ process.
In the rapidity spectra,
each out-going particle of the $2 \to 3$ process
have distinctive features:
The heavy $T$-quark is produced in the central region with $y \lsim 1$,
while the rapidity spectrum of the $b$ quark is rather flat and
that of the other light quark-jet is of M-shape.
We have also considered the $s$-channel process of $q \bar{q}' \to T \bar{b}$,
which is expected to be sizable
if the resonant decay of the heavy charged gauge boson
$W'$ is possible.
We examined the parameter space for $M_{W'}  > M_T$,
which was found to be consistent with the EWPD.
Unfortunately, the total cross section is too suppressed
due to a very  small region
of the resonant decay.

Finally, we have taken into account of the gluon fusion production of
the heavy $T$-quark and the SM top quark, which is mediated by
the Higgs boson and the pseudo-scalar $\eta$.
This can be substantial when the heavy $T$-quark is  light enough.
However, we found that the total cross section is still
suppressed, only ${\cal O}(10^{-3})$
of the dominant $2 \to 2$ process.
An interesting feature is that the $\eta$ contribution is dominant
over the Higgs contribution for relatively light $M_T$.

\acknowledgments
The work of C.S.K. was supported
in part by  CHEP-SRC Program and
in part by the KRF Grant funded by the Korean Government
(MOEHRD) No. KRF-2005-070-C00030.
The work of JS is supported by KRF under grant No. R04-2004-000-10164-0.
The work of KYL is supported by Korea Research Foundation Grant
(KRF-2003-050-C00003).  The work of KC is supported by
the National Science Council of Taiwan under grant no.
94-2112-M-007-010- and 95-2112-M-007-001-.

\appendix
\section{Helicity amplitudes and kinematics for $q g\to q' \bar{b} T$}

For the process of
\beq
u(p_1) +  g(p_2) \to d (k_1) +  T(k_2) + \bar b (k_3),
\eeq
the amplitude-squared
(after summing over the colors and helicities and averaging over
initial colors and helicities) is
\begin{eqnarray}
\label{eq:23ampsq}
 \overline{|M|^2}
&=& \frac{384}{4 \cdot 24} g_s^2  \\
&\times& \Biggr\{
  \,\left|Q_{p1k1} \right|^2 \,  \biggr [ 2 \, D^2_{k2p2} \langle k_3 p_1\rangle
      \left\{ \langle k_1 p_2\rangle \langle k_2 p_2\rangle
             +M_T^2 \langle k_1p_2\rangle -M_T^2 \langle k_1 k_2\rangle  \right\}
\nonumber \\
 && \hspace{40pt}
  +  \, D^2_{p2k3} \langle k_1 k_2 \rangle \left\{ -2 m_b^2 \langle k_3  p_1 \rangle
                       + \shat  \langle k_3  p_2 \rangle
                       + \shat \,m_b^2  \right\} \nonumber \\
 &&\hspace{40pt}
  +  \,D_{p2k3} \,D_{k2p2} \big\{
      2 \langle k_1  k_3 \rangle \langle k_2  p_2\rangle \langle k_3  p_1 \rangle
     - 2 \langle k_1  k_2 \rangle \langle k_3 p_1\rangle \langle  k_3 p_2\rangle
+ 4 \langle k_1 k_2 \rangle \langle k_2 k_3\rangle \langle k_3  p_1\rangle\no \\
&&\hspace{107pt}
     - 2\langle k_1 k_2 \rangle \langle  k_2 p_2 \rangle \langle k_3 p_1 \rangle
     + 2 \langle k_1  k_2  \rangle \langle  k_2   p_1 \rangle \langle  k_3  p_2 \rangle
     - 2 \langle k_1  p_2  \rangle \langle  k_2   k_3 \rangle \langle k_3  p_1 \rangle
\no \\ &&\hspace{107pt}
     - \shat \langle k_1  k_2 \rangle \langle  k_2  k_3  \rangle \big\}
                \biggr] \nonumber \\
&& +  \left|Q_{k2k3}\right|^2
 \biggr [ 2 D_{k1p2}^2 \langle k_1   p_2 \rangle \langle   k_2  p_2  \rangle \langle   k_3 \rangle  p_1
                        +  D_{p1p2}^2 \,s \langle k_1   k_2  \rangle \langle   k_3   p_2\rangle
                       \nonumber \\
&& \hspace{40pt} +  \,D_{k1p2} \,D_{p1p2} \big\{
     - 2 \langle k_1  p_2  \rangle \langle  k_2   p_1  \rangle \langle  k_3   p_1 \rangle
     - 2 \langle k_1  p_2  \rangle \langle   k_1  k_2  \rangle \langle  k_3  p_1 \rangle
 + 2 \langle k_1  p_1  \rangle \langle   k_2  p_2  \rangle \langle   k_3  p_1 \rangle
\nonumber \\
&& \hspace{110pt}
     - 2  \langle k_1  p_1  \rangle \langle   k_1  k_2  \rangle \langle  k_3  p_2  \rangle
- 4 \langle k_1  p_1 \rangle \langle   k_1  k_2  \rangle \langle  k_3  p_1\rangle
\nonumber\\
&& \hspace{110pt}
     + \shat \,\langle k_1  k_2  \rangle \langle   k_3  p_1\rangle
+   \shat\, \langle k_1  k_2  \rangle \langle   k_1  k_3 \rangle
\big\}
    \biggr ]
 \Biggr\}
\end{eqnarray}
where $\langle p_i p_j \rangle \equiv p_i \cdot p_j$,
and the propagator factors are
\begin{eqnarray}
 \shat &=&  (p_1 + p_2)^2, \quad
D_{k1p2} =  \frac{1}{ (k_1 - p_2)^2 } \nonumber \\
D_{p1p2} &=& \frac{1}{ (p_1 + p_2)^2 }, \quad
D_{k2p2} = \frac{1}{ (k_2 - p_2)^2 - m_T^2},\quad
D_{p2k3} = \frac{1}{ (p_2 - k_3)^2 - m_b^2} , \no \\
Q_{p1k1} &=& \frac{g^{W}_{ud} \, g^{W}_{Tb}}{ ( p_1 - k_1)^2 - m_w^2 }
          + \frac{g^{W'}_{ud} \, g^{W'}_{Tb} }{ ( p_1 - k_1)^2 - M_{W'}^2 },
  \nonumber \\
Q_{k2k3} &=& \frac{g^{W}_{ud} \, g^{W}_{Tb}}{ ( k_2 + k_3)^2 - m_w^2 }
          + \frac{g^{W'}_{ud} \, g^{W'}_{Tb}}{ ( k_2 + k_3)^2 - M_{W'}^2 }.
  \nonumber
\end{eqnarray}

In order to specify the kinematic variable we define
the energies of involving particles, in the parton c.m. frame, by
\bea
E_u &=& \frac{\sqrt{\shat}}{2}, \quad E_g = \frac{\sqrt{\shat}}{2},
\\ \no
E_d &=& \frac{\sqrt{\shat}}{2}(1-\eta), \quad
E_\bbar = \frac{\sqrt{\shat}}{2}(1-\zeta), \quad
E_T =\frac{\sqrt{\shat}}{2}(\eta+\zeta)
\,.
\eea
Then the final three momenta ($\vec{p}_d, ~\vec{p}_\bbar,~\vec{p_T}$)
lie in a plane.
We define the angle $\theta$ by
\beq
\cos\theta = - \hat{p}_d\cdot \hat{p}_\bbar =
1-2\,\frac{\eta\zeta-M_T^2/\shat}{(1-\eta)(1-\zeta)}
\,.
\eeq
Kinematic boundaries for $\cos\theta=\pm 1$ lead to the
allowed parameter space of
\beq
\label{eq:Dalitz}
\eta\,\zeta \geq \frac{M_T^2}{\shat},\quad \eta+\zeta \leq 1+\mu_T
\,.
\eeq
And $\alpha \in [0,2\pi]$ and $\cos\beta \in [0,1]$.

\end{document}